\documentstyle[epsf,aps,pre]{revtex}
\title{A self-similar model for shear flows in dense granular materials}
\author{G. Debregeas\thanks{present address: Institut Charles Sadron, CNRS UPR 22, 6, rue Boussingault, 67 083 Strasbourg C\'edex} and C. Josserand
\thanks{present address: Laboratoire de Mod\'elisation en M\'ecanique,
CNRS UMR 7607, Universit\'e Pierre et Marie Curie, Case 162, 8 Rue du 
Capitaine Scott, 75015 Paris-France}\\
\small\it The James Franck Institute, The University of Chicago,\\
5640 South Ellis Avenue,\\
Chicago, Illinois 60637, USA
}

\begin{document}

\maketitle

\begin{abstract}
 We propose a model to
describe the quasistatic shearing of dry granular materials, which notably
captures the differences in velocity profiles recently observed in 2
and 3-D Couette flow experiments. In our scheme, the steady-state flow
is due to the intermittent motion of particle clusters moving
together with the wall. The motion of a cluster is associated with the
transient formation of a fracture inside the sheared pack. The model is based on
the existence of a persistence length for the fractures, which imposes a 
self-similar structure on the clusters.
 Through a probabilistic approach, we can evaluate the rate of
 appearance of a cluster of a given size and obtain a prediction for
 the average velocity profiles. We also predict the
 existence of large stress fluctuations at the moving wall, which
 characteristics are in good agreement with experimental data.
 \end{abstract}

\section{Introduction} 

Dry granular materials exhibit a large range of dynamical behaviours (for a 
general review see
\cite{JNB}). An assembly of non-cohesive macroscopic particles confined in a 
box can be considered as a solid, but will start   
flowing when submitted to a large enough shear stress. However, the underlying
mechanism can be very different, depending on the shearing rate imposed and the
resulting density of the material. Upon rapid shearing or shaking, a fluidized
 state is
reached in which the stress is transfered from the boundaries to the bulk
through binary collisions. This dynamics is well understood
under the scope of the so-called kinetic theory \cite{Savage}. 
In many practical situations however,
gravity maintains a high density in the material so that the stress is mainly 
transfered by rubbing friction between particles in persistent contact. In 
this regime, all relative motion between particles tend to be confined to 
very narrow regions of the material, making a continuous approach inadequate. 
Within these so-called shear band, each particle is submitted to large forces
which rapidly fluctuate as the packing
environment and the associated force network evolve.
The flow is driven by collective and jerky moves of large sets of particles, 
hold together by transient force chains. Although this regime is relevant
to soil mechanics and geology (earthquakes and pyroclastic flows for
example\cite{straub}) and important in industrial applications
(granular transport in hoppers and pipes), a satisfactory description
is to date lacking for such fully developed dense flows\cite{Jackson}. 

A strong experimental effort has however been seen recently to probe
 the grain-scale dynamics within these shear bands, which now provides a good 
test for physical models: two series of experiments have been performed 
independently on 2-D and 3-D Couette cells. A collection of particles is confined between a fixed outer cylinder and a rotating inner one. In the 2-D experiment, the system consists of one monolayer of disc-like particles squeezed between 2 horizontal plates. In this geometry, a shear 
band is always present at the vicinity of the rotating inner cylinder. By different 
techniques, the authors were able to precisely measure the decay of the 
average particle velocity from the inner moving wall to the immobile region 
towards the outer cylinder\cite{behr2d,mri}.
 Surprisingly enough, the two results were found to
be significantly different: in 2-D the velocity $V$ decayed with the
distance $r$ to the inner  wall according to an exponential law,
whereas in 3-D, the velocity profile was well fitted by 
a Gaussian centered on the surface of the wall. These results showed to be robust for
polydisperse or highly irregularly shaped particles. For round and monodisperse
grains however, a layering effect and a strong slippage between adjacent layers were observed 
 which result in more complicated profiles.

In this note we will focus on the Gaussian/exponential behaviours as the 
spatial dimension changes. Because a Gaussian form cannot be derived 
from a local differential equation (for example see \cite{Bouchaud}), 
the 3-D velocity profile suggests that a purely local model would fail to describe it. 
Here we introduce non-local effects by postulating a correlation length in the particles 
displacements which depends on their position within the shear band. 
 This approach comes to modeling the flow as a succession of transient fractures allowing 
 the coherent motion of clusters of various sizes.
 This hypothesis alone allows us to produce an average velocity profile consistent with what
was observed in both 2- and 3-D. Finally, we will show that the probability
distribution and the force spectrum derived from this model are in good
agreement with experimental results found in the literature.

\section{The model}

\begin{figure}[schema]
\centerline{ \epsfxsize=14truecm \epsfbox{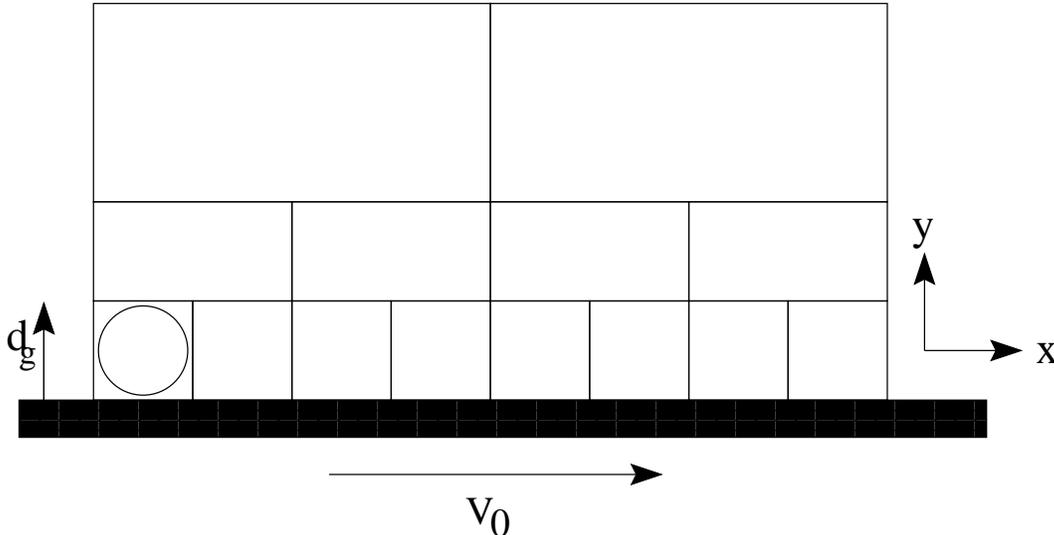} }
\caption{\protect\small Self-similar construction describing the shear flow. 
The wall moves at a constant velocity $V_0$ as well as the first layer of 
particles. At a distance $r$ from the wall, the flow is due to the sudden 
motion of clusters of size $r$ (or bigger) in all directions. The lines 
represent the possible location of fractures that appear in the deforming 
pack during the shearing process. 
\label{schema}}
\end{figure}

We consider a plane shear geometry: a wall is moving at a constant speed $V_0$
 along a half-space of granular material. The flow velocity is assumed to 
vanish far from the moving wall (see figure \ref{schema}). We impose 
a no slip condition at the wall so that $V(0)=V_0$ (this boundary condition 
mimics a classical experimental realization where the first layer of 
particles is glued onto the moving wall). In our scheme, the motion of each
particle can be decomposed into successive steps of typical length
$d_g$ occurring at velocity $v=V_0$. Between two steps, the particle remains 
immobile. One grain can thus either be at rest or move together with the 
wall\footnote{note, this bimodal
velocity distribution for the grains have been roughly observed in
experiments\cite{Veje}}. The interface between moving and static
particles defines a failure surface for the pack. We define $P(r)$
as the stationary probability of a failure surface to be present, per unit of 
time and length (in units of particle diameter) along the $y$-axis, at a 
distance $r$ from the moving wall (as in ref \cite{pouliquen}). 
Although a reduction in granular density by roughly $10 \%$ - the so-called 
Reynolds dilatancy - is known to accompany shear flow, we will first consider 
a uniform density. Both two and three dimensional geometries will be 
addressed; $D$ will denote in the following the space dimension. A 'fracture' 
will either be a line in 2-D, or a plane in 3-D.

We first consider the dynamics of the first freely moving layer in the region $d_g<r<2d_g$.
 As the wall moves by a distance $d_g$ in a time $\tau_0=d_g/V_0$, a particle
located in this layer is either dragged by the wall producing a failure
further away, or stays immobile so that a crack develops at exactly $r=d_g$.
Assuming a Coulomb-type slip condition, the probability of the latter
 depends on the ratio of
the normal and shear stress to which is submitted the particle
during the time period $\tau_0$. Based on this mechanical
consideration, and by analogy with thermally activated processes,
Pouliquen \& al\cite{pouliquen} have proposed an expression for the
probability $p_0$ of slippage between $2$ particles from $2$
consecutive layers in a granular material slowly flowing down a pipe.
In their picture, the spatial stress fluctuation induced by the
randomness of the packing structure plays the role of a temperature, allowing 
them to apply classical rate processes theory\cite{eyring}. In the quasi-static
regime we address here, the average shear and normal stress are uniform 
in the material since each layer is constantly in mechanical equilibrium.
In the limit of uniform density, it is therefore natural 
to assume a uniform value for $p_0$.
Although, the derivation of the slippage rate not only requires the knowledge 
of the static probability $p_0$. A typical fluctuation time-scale, related to 
the flow itself also needs to be introduced.
This time-scale corresponds to the relaxation time of the stress network around the chosen particle. 
For a particle in the first layer, the only relevant time-scale is $\tau_0$ which is the time needed for a shear
strain of $1$ to be established. We can now write down the rate of
slippage at the wall $P(d_g)$ as :

\begin{equation}
 P(d_g)\cdot dt\cdot \frac{dr}{d_g} = {\cal A} \cdot p_0 \cdot \frac{V_0 \cdot dt}{d_g}\cdot \frac{dr}{d_g} 
\label{grain}
\end{equation}

${\cal A}$ is a constant introduced for normalization. 
It should be noted that we implicitly suppose a clear separation between 
the successive hopping events or equivalently that $p_0<<1$ .

The calculation of $P(r)$ away from the wall is based on a self-similar
argument. We postulate that, for a yielding event to occur at a
distance $r$ from the wall, the crack must extend radially over a
length of the order $r$. Hence the shearing of the material at a
distance $r$ from the wall requires the coherent motion 
of a solid cluster of size $r$ in all
directions. This cluster is dragged by the wall over a distance $d_g$ as one solid object,
in the same way as was the single particle from the first layer (see figure \ref{schema}). 
This self-similar description of the cluster size is introduced to account 
for the perturbation of the force network in the vicinity of the wall. 
Chain forces, which are responsible for the rigidity of the dense pack,
 are screened by the proximity of a solid boundary. Independent motion of particles are thus only allowed
in the vicinity of the wall whereas the bulk behaves as a solid body.
 We note that an identical scaling have 
been successfully postulated in (at least) 2 very different contexts to
describe the effect of a wall on dynamical structures. To 
estimate the Prandtl mixing layer in turbulent flows near a wall, the 
characteristic length of the vortices are assumed to increase like the 
distance to the boundary layer. The derivation of polymer dynamics near 
a wall in the semi-dilute regime also assumes that the `blob' size (the 
dynamical coherence length of the monomers) grows like the distance to the 
wall\cite{poly}. As the width of the shear band in granular flows is always 
of the order of $10$ bead diameters, regardless of the bead size, the bead 
diameter appears to be the only relevant length-scale in this problem. A 
linear increase of the coherence length with the distance to the wall is 
therefore the only reasonable choice\footnote{A different mapping of the 
coherence length $l(r)$ may have to be used for a different 
geometry. For example, the stationary dense granular flow along 
an inclined plane seems to exhibit a characteristic length $H$
depending on the inclination and roughness of the inclined 
plane\cite{mills,pouliquen2}. We postpone the discussion of the coherence 
length in this case for a further study.}. For simplicity we will take 
exactly $r$ as the coherence length in the following.

Given this general picture, the scaling analysis follows: we define $N_g(r)$
 as the number of particles involved in a crack developing at a distance $r$ : 
$$ N_g=\left(\frac{r}{d_g}\right)^{D-1}$$

The probability for this set of particles to become simultaneously unstable is $(p_0)^{N_g}$, 
whereas the time-scale $\tau(r)$ of the stress fluctuations experienced by the cluster
of size $r$ reads:
$$ \frac{1}{\tau(r)}=\frac{V_0}{r} $$

The scaling for $\tau(r)$ was made consistent with the criterion used
for the first layer: $\tau(r)$ corresponds to a shear strain of 1 for
the cluster of size $r$. Two different time-scales are thus involved in
the dynamics of these blocks: a rapid one, $\tau_0$ associated
with the release of the stress when a yielding occurs; a slower one, $\tau(r)$ 
which characterizes the stress fluctuations experienced by the particles at a distance $r$ from the wall.

Finally, a particle located at $y=r$ has $N_g$ possible locations along
the crack and the expression for $P(r)$ eventually reads:

\begin{equation}
 P(r)dt \cdot \frac{dr}{d_g} = A \cdot N_g \cdot (p_0)^{N_g} \cdot
\frac{V_0 dt}{r} \cdot \frac{dr}{d_g}
\label{proba}
\end{equation}

When yielding occurs at a distance $r$ from the wall, every particle below 
$y=r$ moves over a distance $d_g$ whereas no motion occurs above $y=r$. Such
an event represents a discontinuity in the instantaneous velocity
profile (the velocity is $V_0$ below the crack and $0$ above during a
time $\tau_0$). The constitutive differential equation for the
stationary mean velocity profile $V(r)$ thus reads :

\begin{equation}
 \frac{\partial V}{\partial r}=- P(r)
\label{eqdif}
\end{equation}

with the boundary conditions $V(0)=V_0$ and $V(\infty)=0$..

For $D=2$, we obtain:
$$ \frac{\partial V_{2D}}{\partial r}=-\frac{A}{d_g} V_0 \left(p_0
\right)^{r/d_g} $$

which by integration gives an exponential velocity profile $V_{2D}(r)$:

$$ V_{2D}=V_0 \cdot e^{-\frac{r}{\lambda d_g}} \quad {\rm with} \quad
\lambda= -\frac{1}{{\rm ln}(p_0)} \quad {\rm and} \quad A=\frac{1}{\lambda} $$

For $D=3$, it gives:
$$ \frac{\partial V_{3D}}{\partial r}=-\frac{A r}{d_g^2} V_0 \left(p_0
\right)^{(r/d_g)^2} $$

so that the velocity profile $V_{3D}(r)$ is Gaussian:

$$ V_{3D}=V_0 \cdot e^{-\frac{r^2}{2 (\sigma d_g)^2}} \quad  {\rm with}
\quad \sigma^2=-\frac{1}{2{\rm ln}(p_0)} \quad {\rm and} \quad
A=\frac{1}{\sigma^2} $$

The function $P(r)$ then reads respectively in two and three dimensions:

$$ P_{2D}(r,t)=\frac{V_0}{\lambda d_g}e^{-\frac{r}{\lambda d_g}} \quad {\rm and} \quad P_{3D}(r,t)=\frac{r \cdot V_0}{\sigma^2 d_g^2}e^{-\frac{r^2}{2(\sigma d_g)^2}}$$

Although a few hypothesis have been used to derive these results, one should note
that the major assumption is that the persistence length of the
fractures increases linearly with $r$. This argument alone produces the
main behaviour of $P(r)$ as a function of $p_0$.  The rescaling of the
stress fluctuation time-scale $\tau(r)$ only controls the pre-factor of
the exponential and Gaussian velocity profiles. A correction could also 
arise from a spatial dependence of $p_0$. For instance, the shear induced density
profile neglected in the present approach might affect the uniformity 
of the average shear and normal stress per particle within the shear band.
Considering that $p_0$ is a decreasing function of the density,
this correction may increase the velocity decay and narrow the shear band.  
This may explain deviations from the Gaussian and exponential profiles
observed in experiments with rounded and monodisperse particles where 
the largest density gradients are present. 
 However, an exact solution of (\ref{eqdif}) for
a general $p_0$ is highly non-trivial and should be investigated in a
further work.

\section{Forces distribution and power spectrum}

In addition to the velocity profiles, the knowledge of the function
$P(r)$ gives appropriate information for computing other properties of
the flow such as the probability distribution function (PDF) of the
forces acting on the wall as well as the power spectrum of this
signal.  These two quantities have been experimentally studied in both
two and three dimensional granular shear flows\cite{behr2d,behr3d}. In
the latter experiment, the normal force on the bottom of a shear cell
was monitored using a force transducer which could  accommodate a large
number (from 4 to 100) of particles. The resulting measurement
corresponded to the integrated value of many individual contact forces.
The measured force signal F(t) appeared as a series of narrow peaks of
various heights. To compare these results with the present model, we suppose 
that each peak is the signature of a failure in the structure of the sheared 
pack. Assuming that the slippage of one grain releases a given force, the 
motion of a cluster of size $r$ produces a peak of intensity 
$F(r)$ proportional to $N_g$, or equivalently:

$$ F(r) \propto r^{D-1} $$

These peaks last for a time $\tau_0$ and are uncorrelated in time.
>From here, we can derive the form of the probability distribution function 
(PDF) $\rho(F)$ by identifying $r^{D-1}$ with $F$ in $P(r)$:

\begin{equation}
 \rho_{2D}(F) = \rho_{3D}(F) = \frac{{\rm exp}(-F/F_0)}{F_0} 
\label{densite}
\end{equation}

where $F_0$ is the mean force of the distribution. Surprisingly enough, these
PDF's are the same for 2- and 3-D. Both agree with experimental observations
at large forces\cite{behr2d}. The behaviour at low forces is not probed by 
experiments although it may have more complicated features.

One may note that eq. (\ref{densite}) is formally identical to what
has been observed for large forces on individual
particles in a compressed static array\cite{dan}. But this
static distribution would narrow upon spatial averaging. By contrast,
 the width of the force distribution in the present description is insensitive
 to the integration area\footnote{The finite size should merely show up as a
cut of at high forces.} in agreement with experimental observations. In
the continuously sheared regime, the wide force distribution arises
from the coherence in the force release induced by the sudden motion of
large clusters, and not from the purely static force distribution
on individual particles.

The power spectrum of the force fluctuations can also be obtained in
this description as follows: the power signal is decomposed into a
stochastic succession of peaks of height $F^2(r)$ and width $\tau_0$.
The period $\tau'(r)$ between two successive pulses of amplitude $F^2(r)$ is 
simply the inverse of the rate $P(r)$. For a given $r$, and therefore a given 
$F(r)$, the signal is the so-called telegraph noise, for which the power 
spectrum $Q_r(\omega)$ reads\cite{machlup}:

$$ Q_r(\omega)=\frac{F^2(r)}{\pi} \frac{\tau'(r) \tau_0}{(\tau'(r)+ \tau_0)^2}
\cdot \frac{1/T(r)}{\omega^2+(1/T(r))^2} $$
$$ {\rm with} \quad 1/T(r)=1/\tau'(r)+1/\tau_0 $$

Therefore, since failures at different distances from the wall are 
uncorrelated, the power spectrum $Q(\omega)$ for the overall signal is given 
by the sum:

\begin{equation}
Q(\omega) = \frac{1}{d_g} \int_0^\infty Q_r(\omega) dr  
\label{power}
\end{equation}

\begin{figure}[h]
\centerline{ \epsfxsize=14 truecm \epsfbox{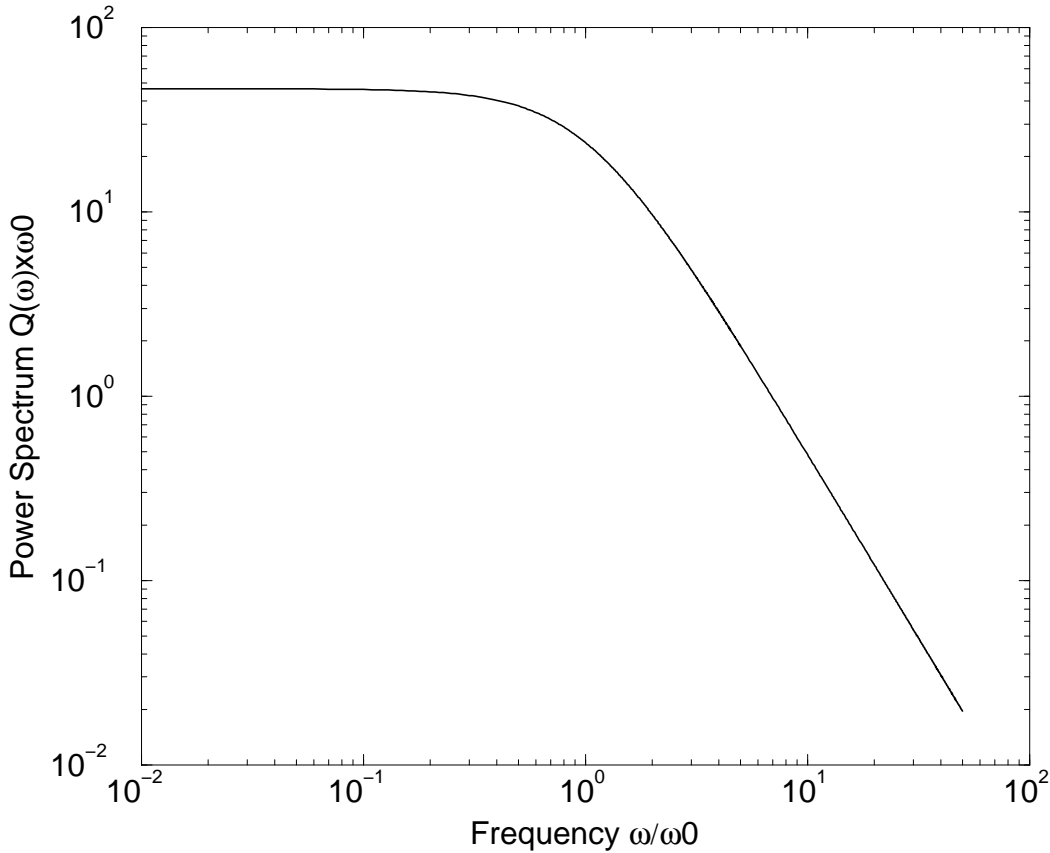} }
\caption{\protect\small Power spectrum of the forces in the 2D geometry. The 
curve has a flat part at low frequencies and a $1/\omega^2$ behaviour for large 
frequencies. The cross-over between these two regimes occurs around the typical 
frequency $\omega_0$ for which the wall has moved by one grain diameter.
\label{spectrum}}
\end{figure}

$Q(\omega)$ exhibits qualitatively the same behaviour for 2 and 3
dimensions, dominated by Lorentzian-like functions (see figure (\ref{spectrum})
). For frequencies much 
larger than $\omega_0=1/\tau_0$, the power spectrum behaves like
$\omega_0/\omega^2$ as observed in experiments\cite{behr3d}; this is
a direct consequence of the stochastic description of the process. On the
other hand, the expression (\ref{power}) displays a well-defined non-zero
limit as $\omega \rightarrow 0$.  By comparison, the power spectra
measured in \cite{behr3d} exhibit a continuous, although small
increase as $\omega \rightarrow 0$.

\section{Conclusion-Acknowledgements}

We have proposed a stochastic description of slowly sheared
granular material that can capture the main characteristics of the flow: the
average velocity profiles and force fluctuations in 2 and 3 spatial dimensions.
We want to underline the relatively good agreement with experimental data
with regards to the small number of ingredients introduced. In
particular, we ignored the existence of granular density variations
along the direction normal to the flow which may induce significant 
corrections. The crucial assumption of this model lies in the self-similar 
structure of moving clusters that rapidly form and disappear as the material
flows. This assumption has been qualitatively suggested by the mere
observation of 3-D Couette flows watched from below through a transparent 
bottom plate where the jerky and multi-scale dynamics is clearly visible. 
We hope that this tentative description will motivate further
quantitative investigations to probe the nature of space and time correlations
 in the motion of neighboring particles in such systems.
 
Here we have focused on recent experimental data in order to test 
this self-similar model but we think that this simple description of dense 
granular dynamics as a super-imposition of coherent moves can be extended
to many more geometries and processes than just plane shear.
The issue is to correctly prescribe the coherence length mapping $l(r)$, and 
describe the probability $P(r)$ for each realization.
In particular, we anticipate that the existence of multiple relaxation times
recently demonstrated in granular compaction 
experiments\cite{christophe&alexis} could be associated with rearrangements 
over different length-scales during the relaxation
of the packing structure. To that extent, this model may offer a clue
to understand the numerous features that granular material share
with glassy liquids below the glass transition (aging, slow relaxation, 
jamming\cite{Liu&Nagel}).

It is a pleasure to thank D. Mueth, H. Jaeger, S. Nagel, T. Witten and
L. Kadanoff for many interesting and stimulating
discussions. G.D. is supported by the David Grainger Fellowship. C.J.
is supported in part by the ONR grant: N00014-96-1-0127, the
MRSEC with the National Science Foundation DMR grant: 9400379, and 
would also like to thank the Argonne National Laboratory for its support.

\end{document}